# Impact of the COVID-19 outbreak on Italy's country reputation and stock market performance: a sentiment analysis approach

Gianpaolo Zammarchi[1], Francesco Mola[1], Claudio Conversano[1*]

**Abstract**

During the recent Coronavirus disease 2019 (COVID-19) outbreak, the microblogging service Twitter has been widely used to share opinions and reactions to events. Italy was one of the first European countries to be severely affected by the outbreak and to establish lockdown and stay-at-home orders, potentially leading to country reputation damage. We resort to sentiment analysis to investigate changes in opinions about Italy reported on Twitter before and after the COVID-19 outbreak. Using different lexicons-based methods, we find a breakpoint corresponding to the date of the first established case of COVID-19 in Italy that causes a relevant change in sentiment scores used as proxy of the country reputation. Next, we demonstrate that sentiment scores about Italy are strongly associated with the levels of the FTSE-MIB index, the Italian Stock Exchange main index, as they serve as early detection signals of changes in the values of FTSE-MIB. Finally, we make a content-based classification of tweets into positive and negative and use two machine learning classifiers to validate the assigned polarity of tweets posted before and after the outbreak.

## 1. Introduction

Twitter is a microblogging and social networking service widely used by users to interact and publish contents in response to events. Twitter is being largely used to share information and express sentiment and concerns on the recent Coronavirus disease 2019 (COVID-19) outbreak, which was first identified in December 2019 in Wuhan, Hubei, China, and resulted in a serious threat to public health worldwide [16]. Italy was one of the first European countries to be severely affected by the outbreak as well as to implement extraordinary measures to limit viral transmission, such as lockdown and stay-at-home orders [30]. This situation might have led to extensive concerns towards Italy, potentially leading to country reputation damage, loss of investments and tourism flows. In particular, since country reputation is usually studied in terms of strategic public diplomacy, effective nation building and nation branding (see, for example, [42]), it is very likely that a dramatic event like the COVID-19 outbreak negatively affects all the three dimensions of the reputation of a country. In the following, we focus specifically on the third dimension (national branding) but concentrate on the reputation of Italy perceived by Twitter's users, the latter measured though sentiment analysis.

Sentiment analysis [26], sometimes referred to as opinion mining or polarity classification, is aimed at analyzing and classifying text into sentiments with a polarity or specific emotions using different approaches. In today's world, communication has become more and more bidirectional, involving not only a company or other entities sending messages/information to people but also vice versa. The widespread use of social networks allows billions of people from around the world to have a direct interaction with companies, actors/actress, sportsmen/sportswomen or politicians, just to give some examples. Having a tool able to understand what a text means without the need of a human interpretation is of huge value, e.g. for a company who needs to reply to thousands of e-mails, letters, messages or any other type of communication. Social media could be a huge source of information, but the sheer number of messages exchanged every day makes it very difficult for any company or organization to extract knowledge from them in a quick way. For example, it would be useful to know that, due to a given event, many people in a specific area of the world need something. We might add that this information could be useful in different situations, e.g. a company trying to intercept the unexpressed needs of potential customers but also a humanitarian organization willing to help people in areas struck by war or natural disasters. Although these two situations can be thought as very distant cases, they share the importance of correctly interpreting a message. This is a very hard task and sentiment analysis tackles only a part of it, as it is focused on polarity and strength of a text. Moreover, sentiment analysis usually has two (positive, negative) or three (positive, neutral, negative) possible outcomes and can be performed at three different levels: document level, sentence level and aspect level. At a document level, a document can be thought as a collection of sentences (e.g. a product review or a book chapter). Each sentence is evaluated separately and then aggregated to express an overall sentiment. At a sentence level, the evaluation is made for every single sentence, resulting in a more fine-grained analysis than document level. Finally, at an aspect level, the analysis can be performed for every single aspect of interest. For example, in a product review we might be interested in price, package, delivery and many other aspects. Every aspect is independent since we can have a positive sentiment for price and a negative sentiment for package in the same product review.

There are two main approaches to perform sentiment analysis: a lexicon-based approach and a machine learning approach. The former uses a list of words (lexicon) associated with a specific sentiment polarity. The sentiment of some document, sentence or text in general is given by the number and strength of positively/negatively associated word in that text [40]. It is possible to use an existing lexicon or to build one. The machine learning approach can instead be thought as building a model to accomplish a supervised classification task. Several models/classifiers can be used, e.g. naïve Bayes (NB), Support Vector Machine (SVM), logistic regression, decision trees, neural networks, etc. They work differently and might have different performance, but they all share the same steps: collect, label and process data, split dataset into training and test set, build the model on the training set, assess model performance.

In the above-described framework, the main contributions of this study are summarized as follows:

1) Analyze the temporal evolution of the sentiment towards Italy before and during the COVID-19 outbreak through a sentiment analysis on tweets. Our goal is to show that right after media started spreading the news of the first Italian case of COVID-19 the sentiment towards Italy and the perceived country reputation, in particular the national brand, began to drop significantly. We show, at the end of Section 3.2, that whilst all negative emotions increase, the positive ones have different behavior, linked to the specific type of emotion considered.

2) Compare the changes in tweets' sentiment with the evolution of stock exchange levels corresponding to the prices of the main Italian stock exchange index (FTSE-MIB). We follow the line of research investigating about the ability of sentiment analysis to predict stock market movements (see, for example, [25]) and assume the polarity of a sentiment expressed on Twitter towards Italy can reverberate on several aspects of the life of a country, including the performance of the stock market which is known to be affected by exogenous factors.

3) Evaluate the performance of different machine learning classifiers to validate the polarity of tweets posted before and after the COVID-19 outbreak. We use two of the most popular models (NB and SVM) to perform the task of validating a previously-made content-based classification of tweets into positive or negative. We assume that important differences in classification accuracy observed in the two periods further support the idea of important changes in the contents of the tweets as tweets posted in the period following the outbreak are more concentrated on topics or sentiments related to the ongoing spread of the pandemic as well as on the negative consequences it generates for the country.

The rest of the paper is organized as follows. Section 2 introduces previous studies performing sentiment analysis on tweets in response to real-world important events. Section 3 describes lexicon-based sentiment analysis to identify shifts in the sentiment towards Italy in the different phases of the COVID-19 outbreak. In Section 4, shifts in sentiment polarity are compared with changes in stock exchange market performance. Performances of two machine learning classifiers in the classification of tweets are compared in Section 5 and the conclusions of the study are presented in Section 6.

## 2. Literature review

A number of studies perform sentiment analysis on Twitter data in response to important events. This kind of analysis is considered a good way to measure public opinion since users are free to express their thoughts about any topic having a (potentially) large audience (see, for example [27, 38]). Here, we focus on those related to important events (usually negative) like migrant crisis, terrorist attacks and disease outbreaks.

In 2015, about one million refugees and migrants came to Europe due to war, disease, terrorism, natural disasters as the most frequent causes [41]. Most people first arrived in south Europe (e.g. Italy or Greece, closest to the northern African costs) and then moved to northern European countries (e.g. Germany). Backfried and Shalunts [2] observe shifts in public perception and media coverage with respect to the growth in number of migrants arriving in Europe. Pope and Griffith [28] collect data in English and German to assess if any difference in sentiment could be found comparing tweets in people speaking the two languages. Öztürk and Ayvaz [24] use a similar approach comparing approximately 350,000 tweets in English and Turkish during the Syrian refugee crisis. Authors report that people in Turkey were more concerned about what was happening near their country, while English-speaking people paid more attention to the political implications of the event.

A number of studies use different methods, including machine learning classifiers, to analyze Twitter data with respect to public opinions and concerns on terrorism attacks. The study from Cheong and Lee [5] is one of the first to analyze tweets in a terrorism-related scenario. Given the large number of users and the ease of sharing texts, pictures and videos, people might use Twitter to share news about the attack, ask for help, etc. For example, during the Mumbai attacks in 2008 or the Jakarta attacks in 2009, the news was first reported by Twitter users [5]. The authors of the referred study propose an approach comprising four phases: a) scanning latest trends in Twitter to identify topics to be further inspected, b) collect as many tweets as possible about that topic, c) extract knowledge from tweets collected in the previous phase and d) summarize and visualize the information through reports.

Simon et al. [34] consider on the attack at the Westgate Mall in Kenya, focusing on the importance of a coordinated system to avoid having multiple information sources (e.g. too many different hashtags for the attack or responders), while Burnap et al. [4] analyze how information spreads in a social network after a terroristic attack, using the Woolwick terrorist attack as a case study. They collect more than 400,000 tweets and use a negative binomial regression method to study the dimension and survival time of information related to the attack, showing that tweet sentiment is able to predict dimension and survival time of the information flow [4]. Ashcroft et al. [1] train a machine learning classifier to predict whether or not a message is supporting jihadists groups. Since many terroristic groups (e.g. ISIS) use Twitter to communicate and spread propaganda, it would be very useful to stop this flow at early stages. They use three types of features (stylometric, time-based and sentiment-based) to classify tweets using different methods (SVM, NB and Adaboost). They present promising results supporting the usefulness of an automated approach to aid analyst to detect radical contents on social networks. Güneyli et al. [14] collect tweets from six Turkish political leaders in 2015. In this case study, the focus is on the presence of the terrorism topic in tweets and the attitude of Turkish leaders towards terrorism is described, which results to be the most discussed topic during the election campaign. Garg et al. [12] focus on the Uri (India) terroristic attack in 2016. Almost 60,000 tweets are collected in a one-month period using different hashtags. Tweet's sentiment is analyzed with SVM and NB classifiers and data are further analyzed in terms of reach and retweets, showing how negative tweets have a longer surviving time than positive ones, even if they are reduced in number. Harb et al. [15] examine the emotional response to two terrorist attacks occurred in the UK (Manchester and London) in 2017. They collect and annotate a few hundred tweets as regard to four negative emotions (anger, fear, sadness and disgust) plus surprise and a residual category (none), and use two deep learning models [Convolutional Neural Network (CNN) and Long Short-Term Memory Network (LSTM)] trained with different datasets (e.g. an existing pre-labeled dataset, a dataset automatically labeled using hashtags, etc.). Classifier accuracy is similar for both deep learning models with an F1-score of 63%. Conde-Cespedes et al. [8] try to solve the problem of the identification of accounts violating Twitter rules. Having an automated system to accomplish such a task would be a great benefit in identifying potential threats. Approximately, 200,000 tweets in different languages (mostly Arabic tweets translated in English) are collected before and after the Paris terroristic attacks in 2015. These tweets are classified into two categories: pro-ISIS or neutral. Using a combination of keywords and sentiment analysis scores, SVM is able to reach over 90% of accuracy.

Finally, several studies focus on the analysis of tweets in response to disease outbreaks. Chunara et al. [7] analyze data from news media, Twitter and official reports by the government during the first 100 days of the 2010 Haitian cholera outbreak. This study shows how Twitter and other sources can be used complementary to data provided by government or health institutions. Specifically, trends in volume of informal sources are significantly correlated with official case data and are available up to two weeks earlier, thus being potentially useful to provide timely estimates of outbreaks dynamics. Internet represents one of the major sources used by people to obtain information during the H1N1 influenza virus outbreak [19]. In 2010, Chew and Eysenbach [6] collect about two million tweets related to this outbreak and show that sentiment analysis performed on Twitter data is a valid tool to measure public perception, allowing health authorities to address real as well as perceived concerns. This is particularly important considering that misinformation might allow a disease to spread more quickly, with a significant cost in health and human lives, while correct information might contribute to adopt behavioral changes (e.g. social distancing), especially in the initial stage of an outbreak when a vaccine is not available. Signorini et al. [33] show other ways in which tweets analysis is useful during the outbreak. Specifically, the extraction of information from a live stream of tweets allows to early detect hot spots. Importantly, Szomszor et al. [37] collect about three million tweets and use them to detect trends of infection spreading one week earlier compared to the official reports, showing how Twitter data analysis is useful as an early warning detection system. Smith et al. [35] analyze tweets posted during the flu season in US in 2012/2013. They use machine learning models to separate tweets about flu awareness from tweets about the infection, demonstrating that these two types of tweets show different trends. For instance, they document that levels of awareness drop after the peak, even when infection levels are still high. Using a different approach, Broniatowski et al. [3] analyze Twitter data posted during the same flu season to build a model able to distinguish between tweets reporting an infection from genetic tweets mentioning flu. Using this tool, they are able to predict changes of influenza prevalence with an accuracy of 85%.

From 2014, several studies used Twitter data to analyze reactions to the Ebola outbreak in Africa. A number of these articles focus on the perception of the disease among people living in Western countries (e.g. US). Fung et al. [11] show that, despite the disease outbreak was far away from the US (only few cases hit the US, while the large majority of cases were in Guinea, Sierra Leone and Liberia), people were very concerned about their own safety. This is proved by the high levels of anxiety, anger and other negative emotions that were significantly higher than those observed during the influenza outbreak. Lazard et al. [20] use data collected during a Center for Disease Control (CDC) live Twitter event to extrapolate the main topics people were more concerned about. They highlighted eight topics similar to the ones discussed nowadays by authorities and everyday users during the COVID-19 outbreak (e.g. ways of transmission, symptoms, survival of the virus outside the body and protections). Towers et al. [39] extract Twitter data and web searches from the first days in which the media reported some Ebola cases in the United States (September 2014) until the end of October. They find a strong relationship between video and news related to Ebola and searches performed on Twitter and Google (e.g. Ebola symptoms). Other papers focus on what was happening in Africa. Oyeyemi et al. [23] collect tweets using keywords such as "Ebola" and "prevention" or "cure" and find that many of them are carrying misleading information. They split their data into correct information, medical misinformation and a generic category for tweets not in these two categories (other). Since some of these remedies are medically questionable (e.g. drink salted water), and the potential audience is very large, the Nigerian government itself decided to respond to this misinformation using Twitter. Guidry et al. [13] study how Twitter and Instagram have been used by CDC, World Health Organization (WHO) and Médecins Sans Frontières (MSF) to educate the public about Ebola. They find that Instagram posts were significantly more likely than tweets to feature contents related to risk perception, e.g. information about adverse outcomes. Another study from Liang et al. [21] compare the relevance of dissemination of information on Twitter from important actors (broadcasting) compared to word of mouth (viral spreading). The analysis is performed on all tweets posted about Ebola in a 14-month period (March 2014 – May 2015) to gain insight on the retweeting patterns. Using these data, authors identify four types of users: influential, hidden influential, disseminator and common user. They highlight the relevance of a broadcasting-type communication, concluding that it would be useful for health authorities to establish a partnership with influential users to communicate more efficiently.

Finally, Twitter was an important means of communication also during the Zika virus outbreak in Central and South America. Fu et al. [10] collect a sample of more than one billion tweets reporting the keyword "Zika", posted in English, Spanish and Portuguese from May 2015 to April 2016. They identify 20 topics that were grouped into five themes (impact and reaction to Zika virus, concern for pregnancy and microcephaly, transmission routes and case reports), and find that user-generated contents play a more relevant role as information channels compared to those of the government authorities, highlighting the need to prevent the proliferation of misleading information.

The rationale for this paper is based on previous literature suggesting that Twitter represents a useful tool to study reactions to epidemic events. This paper makes several contributions to the literature. First, other studies usually use a single method to assess the polarity of a text in a sentiment analysis whilst our approach offers a comparison of different methods (Section 3) and also adds an evaluation of text polarity performed with machine learning classifiers (Section 5). Second, while many studies give more emphasis to graphical representation (e.g. wordcloud) or tables (e.g. most used words), in this study we utilize statistical and machine learning tools focused on sentiment scores' trends rather than on the specific content of tweets. We also analyze individual emotions in order to separate the positive part of a score from the negative one and to better understand how these two components behave. Third, we expand the analysis using economic data from the main Italian stock exchange index to find out if the trend in sentiment towards Italy is related to other aspects (e.g. the economy) of the country. To this regard, we observe a strong relationship between country reputation and stock market performance, with sentiment found to serve as an early detection signal (up to two weeks earlier) for potential effects on the stock exchange index values. Fourth, our work considers a period of time starting from the end of 2019 to give a more comprehensive representation of the evolution of sentiment towards Italy, i.e., it is not strictly limited to the COVID-19 outbreak period. In this way, we effectively show the impact that the COVID-19 outbreak had on sentiment towards Italy.
To the best of our knowledge, the present study is the first one using social media opinions to consider the effects of the COVID-19 outbreak on both the reputation of a country and its economy.

## 3. Impact of the COVID-19 outbreak towards Italy's reputation

### 3.1 Data collection

We use lexicon-based sentiment analysis on tweets to evaluate the temporal evolution of the sentiment towards Italy before and during the COVID-19 outbreak. Approximately 1,000 tweets per day, posted in the period October 2019 – May 2020, in English language and reporting the keyword "Italy" were collected using the Python3 GetOldTweets3 library (https://pypi.org/project/GetOldTweets3/), version 0.0.11. In total, 244,000 tweets were retrieved. After data cleaning, consisting in quality control and removal of tweets for which incomplete content was downloaded, 243,846 tweets are retained and used for further analysis.

### 3.2 Evolution of the sentiment towards Italy before and during the COVID-19 outbreak

Preprocessing of tweets (including removal of punctuation marks, hashtags, mentions and links as well as conversion in lower case letters) was conducted in R [29] version 3.6.3. Six different methods were used to evaluate sentiment of collected tweets: sentimentR [31], vader [17] and four lexicons (nrc, afinn, bing and syuzhet) included in the Syuzhet package [18]. For each method, we consider the mean score for tweets collected in a single day. These values were then plotted to observe the temporal evolution of the sentiment towards Italy (Figure 1a). The mean scores were then standardized to consider the different scales used by each method to report the sentiment of a tweet. (Figure 1b).

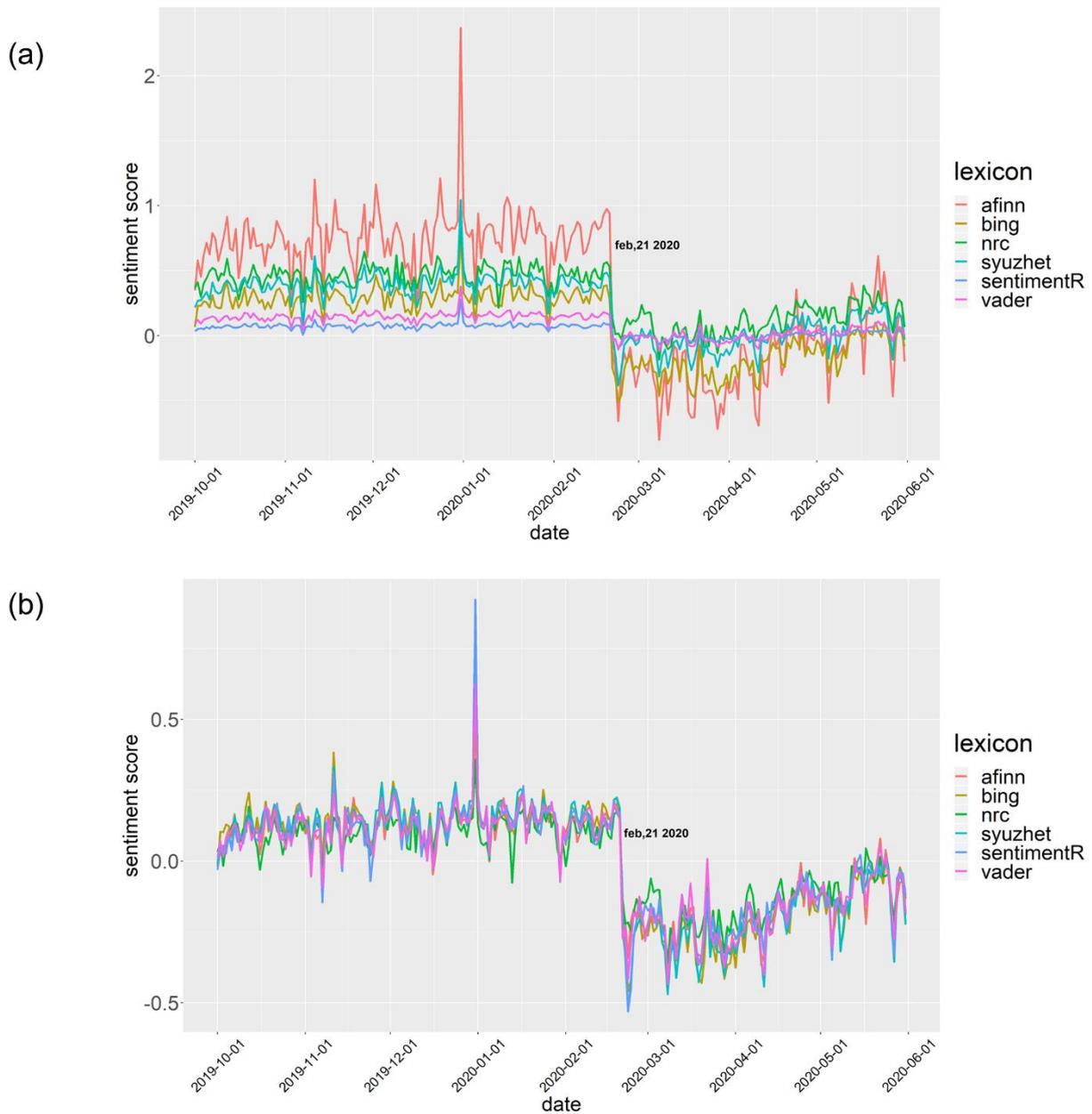

Figure 1. Sentiment score (a) and standardized sentiment score (b) of collected tweets including the keyword "Italy" from October 2019 to May 2020

As shown in Figure 1b, a high concordance among sentiment scores assigned by different methods is observed for standardized sentiment scores. The most positive values, corresponding to the highest peak, are observed on New Year's Eve. Most interestingly, while the trend of sentiment values in the previous months is somewhat stable around neutral or slightly positive values, extremely negative scores are observed from February 21, 2020. At this date, the first Italian case of COVID-19 is reported. From this day, sentiment scores remain negative, although a slow trend towards less negative / neutral values seems to be present in May.

We performed an analysis aimed at finding a structural change in order to verify the existence of a breakpoint in sentiment scores at February 21, 2020. A structural break is a sudden change of values in a time series occurring at one or more specific dates. Therefore, this analysis allows us to gain further insights into the phenomenon object of study by knowing when a significant change occurred in the data. Using the strucchange R package [43] we conduct the structural change analysis for all six used lexicons. For all lexicons, the existence of a breakpoint at 21 February 2020 (observation n. 143), is observed (Table 1). Among lexicons, we notice differences as regards to the optimal number of breakpoints identified according the minimum value of Bayesian Information Criterion (BIC): two breakpoints are identified for bing, nrc and syuzhet (Figure 2) while three are identified for afinn, vader and sentimentR (Figure 3). Additionally, we observe variability in the break dates identified for breakpoints different from February, 21 2020 (Table 1). A graphical representation of the breakpoints obtained from two lexicons (bing and afinn) is reported in Figure 2 and 3, respectively.

**Table 1. Identified break dates using different lexicons**

| Lexicon | Break dates (observation number) | | |
|---|---|---|---|
| Afinn | 82 | 143 | 197 |
| Nrc |  | 143 | 203 |
| bing |  | 143 | 197 |
| syuzhet |  | 143 | 197 |
| vader | 79 | 143 | 198 |
| sentimentR | 83 | 143 | 196 |

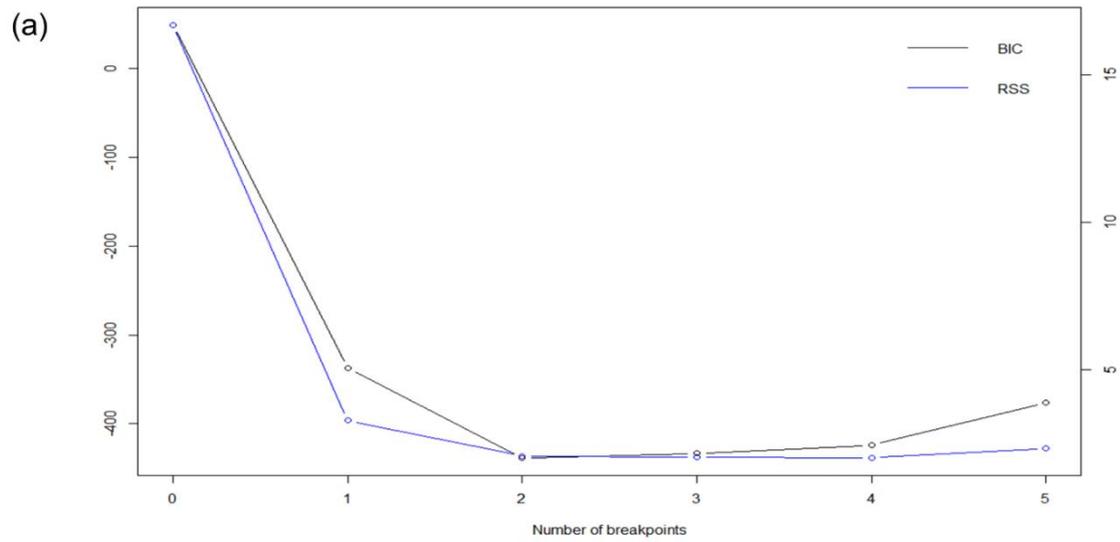

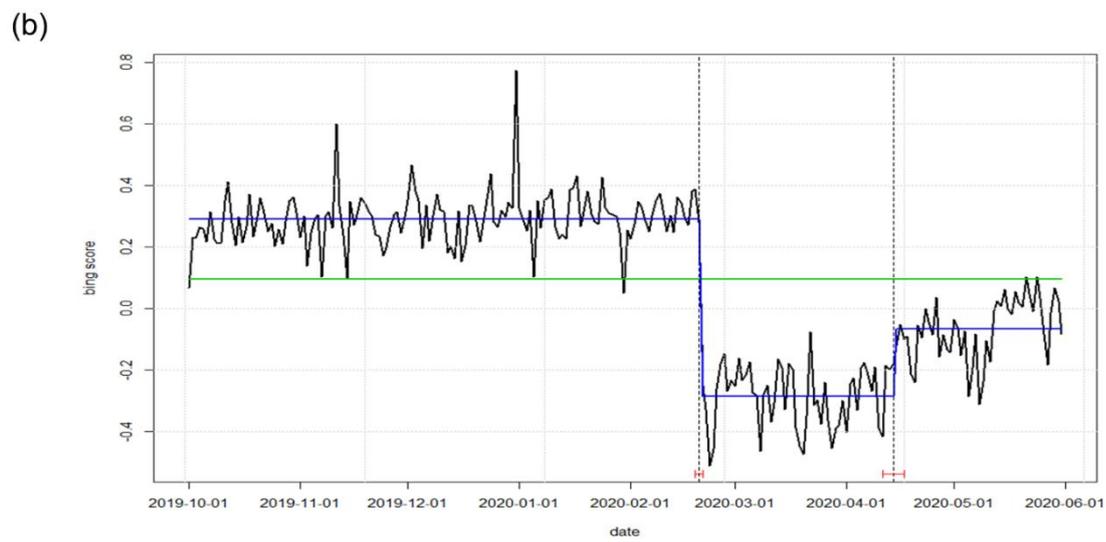

Figure 2. (a) BIC and Residual Sum of Squares using the bing lexicon, (b) Breakpoints in sentiment score using the bing lexicon

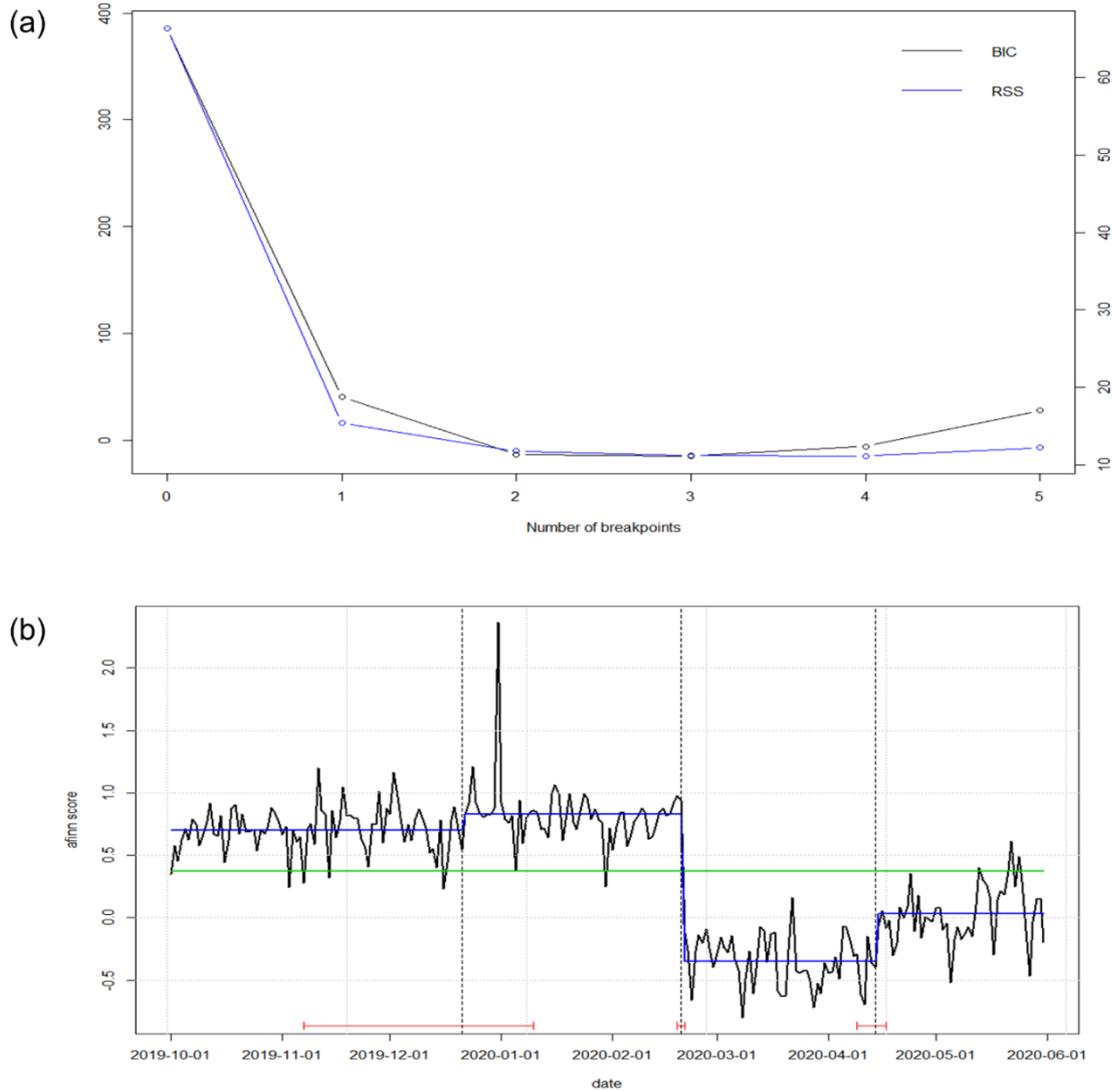

Figure 3. (a) BIC and Residual Sum of Squares using the afinn lexicon; (b) Breakpoints in sentiment score using the afinn lexicon

As we assess that on the date of February 21, 2020 a substantial change of sentiment is observed according to all lexicons, we define two different periods for the subsequent analyses: Period A from October 1, 2019 to February 20, 2020; and Period B from February 21, 2020 to May 31, 2020. Next, we conduct a sub-period analysis of specific positive and negative emotions in these two periods using the nrc lexicon. Besides reporting positive and negative sentiment, this lexicon allows us to evaluate each tweet in terms of eight basic emotions: four negative emotions (anger, disgust, fear and sadness) and four positive ones (anticipation, joy, surprise and trust). The scores for positive and negative sentiment in the two Periods A and B are compared in Supplementary Figure 1, while scores for the eight specific emotions are shown in Supplementary Figure 2. The scores obtained for sentiment as well as for specific emotions show non-normal distribution according to the Shapiro-Wilk test. Moreover, the Mann-Whitney test is used to compare sentiment and emotions between Period A and Period B. As shown in Supplementary Figure 1, general positive sentiment is not decreasing significantly ($p = 0.07$) from Period A to B, whilst negative sentiment is significantly increasing ($p < 0.001$).

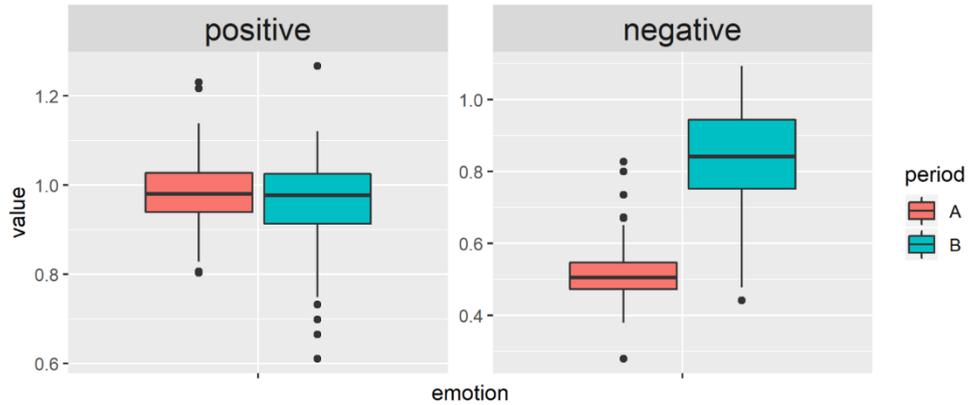

Supplementary Figure 1. Boxplots showing positive and negative sentiment in Period A (October 1 2019 – 20 February 2020) and Period B (21 February – 31 May 2020)

Although a rise of negative emotions in Period B is somewhat expected and in accordance with our hypothesis, positive emotions overall remain stable. This finding is further explored, and partly confirmed, through the analysis of specific emotions. As shown in Supplementary Figure 2, all negative emotions are increasing from Period A to Period B ($p < 0.001$). As for positive emotions, joy is decreasing significantly ($p < 0.001$), whilst anticipation, surprise and trust are increasing significantly ($p < 0.001$). A decrease of joy can be expected during such a hard time, whilst the other three emotions might increase for different reasons. The rising in anticipation and surprise might be interpreted as follows: even if the COVID-19 outbreak is a negative event, it has the power to generate surprise and to increase the desire to know what will happen in the near future. On the other hand, the increase in trust might depend from the willingness to believe in a speedy recovery as well as from the attitude of people living outside of Italy towards encouraging Italians.

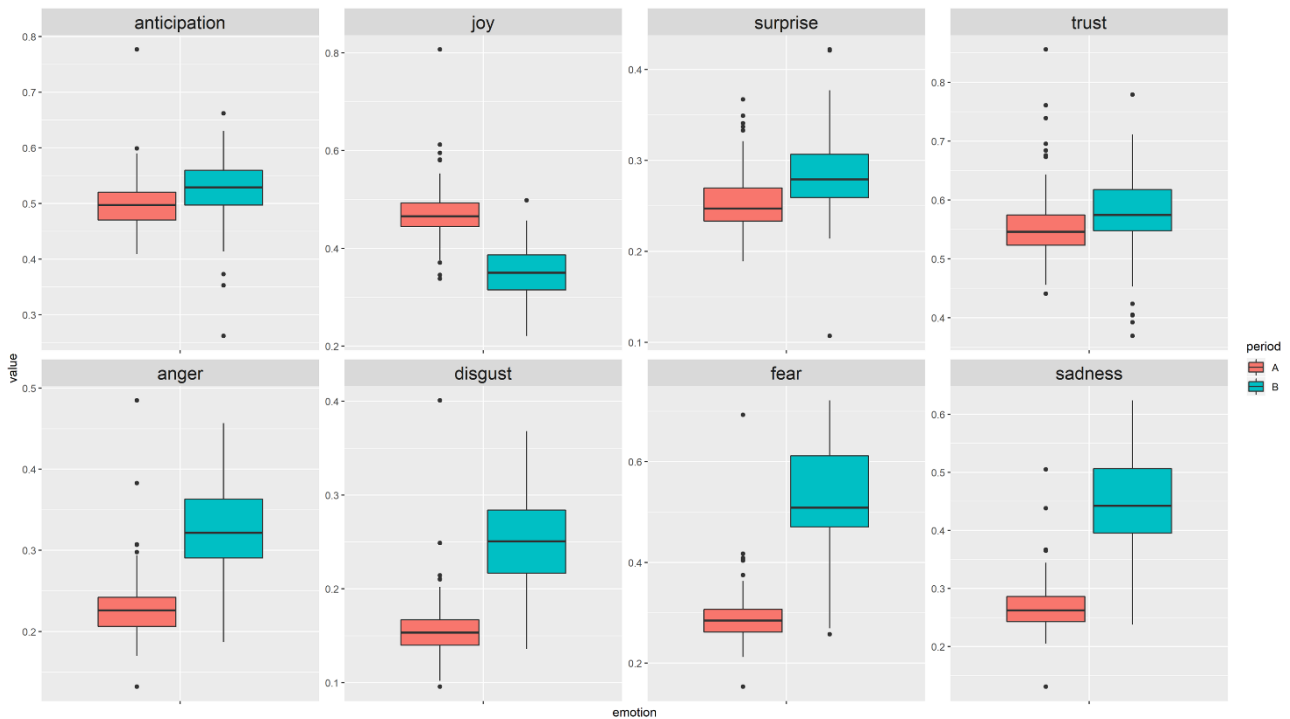

Supplementary Figure 2. Boxplots showing positive and negative emotions in Period A (October 1 2019 – 20 February 2020) and Period B (21 February – 31 May 2020)

Day-to-day detailed temporal evolution for the eight emotions from October 1 to May 31 is shown in Figure 4. We observe that, from February 21, 2020, all negative emotions (right panel) start to rise, whilst positive emotions (left panel) behave differently: surprise rises, joy first decreases and then remains stable, whilst anticipation and trust first decrease and subsequently show a progressive increase in the last days. Figure 4 gives us the chance to grasp subtle nuances not evident in the general trend but which emerge when detailing the various emotions. The rise in the levels of negative emotions is expected since an event such as the outbreak of an epidemic disease certainly has the power to increase anger, disgust, fear and sadness both for those who experience the event firsthand and for those who were not directly affected by it at that time. Positive emotions behave differently but consistently with what could be expected for any specific emotion. For example, joy decreases, and it no longer reaches the values of Period A, while trust has a similar huge drop, but it recovers in a very short time (about a month). This is probably due to the fact that at the time people wanted to express their trust that things would get better for Italy. Slogans as "Andrà tutto bene" (it will be okay) were very popular in Italy during the spread of the pandemic.

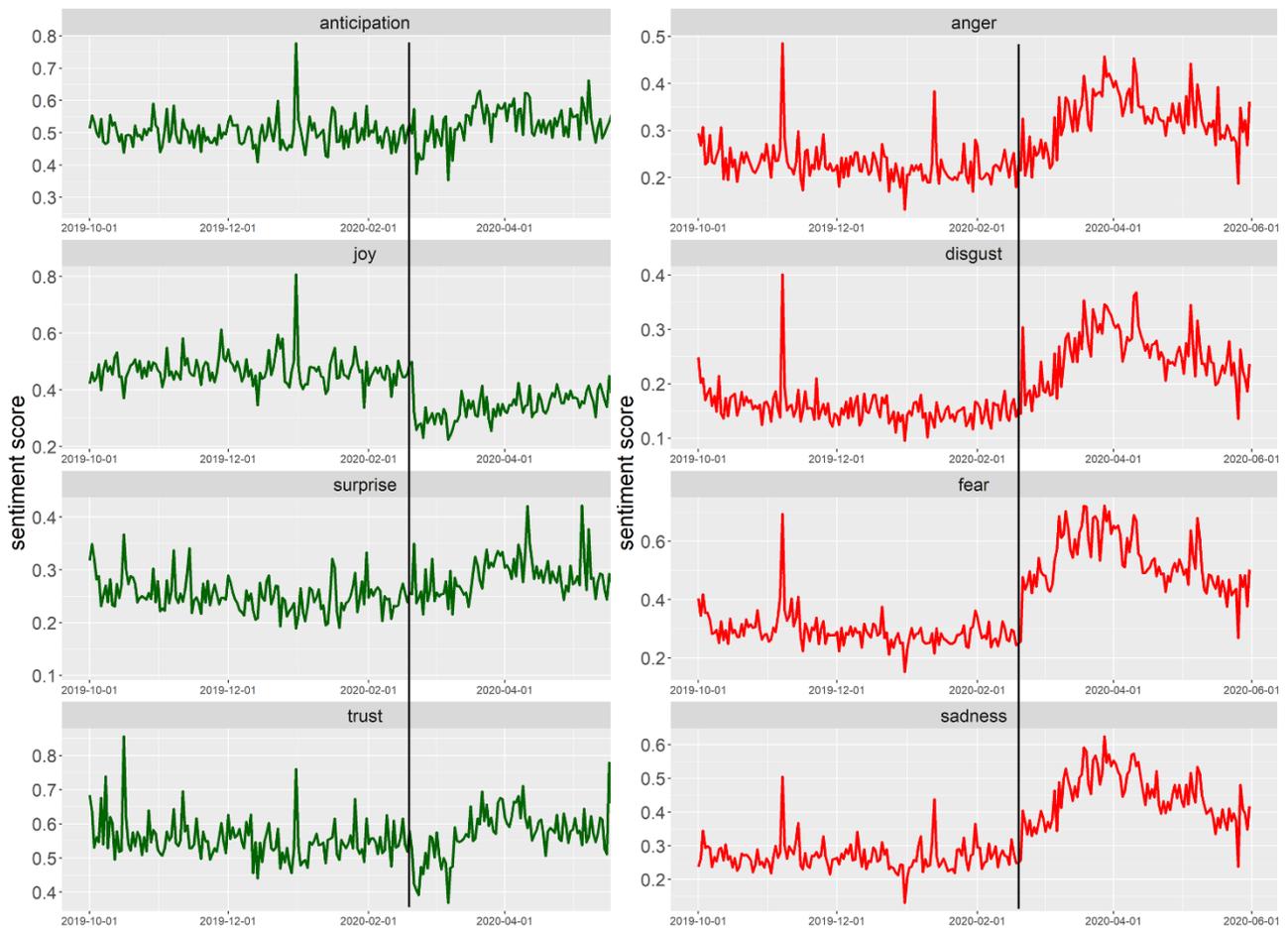

Figure 4. Temporal evolution of the positive (on the left: anticipation, joy, surprise and trust) and negative (on the right: anger, disgust, fear and sadness) emotions from October 1 to May 31.

## 4. Analysis of FTSE-MIB index from October, 1 2019 to May, 31 2020

To verify whether a trend similar to the one observed for the tweets' sentiment is also observed for changes in stock exchange values movements, data from the main Italian stock exchange index (FTSE-MIB) are collected. Closing values of FTSE-MIB are observed from October 1, 2019 to May 31, 2020. Values concerning weekends and other festive days are linearly interpolated to be able to compare homogeneously trends of tweets' sentiments and stock exchange values. The analysis performed on FTSE-MIB evidences the existence of three breakpoints on November 6, 2019, March 7, 2020 and April 26, 2020 (Figure 5). Specifically, the first breakpoint observed during the COVID-19 outbreak period dates March 7, 2020: it happens 15 days after the change in sentiment scores (February 21, 2020). Interestingly, the day after the observed breakpoint the Lombardy region was set into lockdown.

(a)

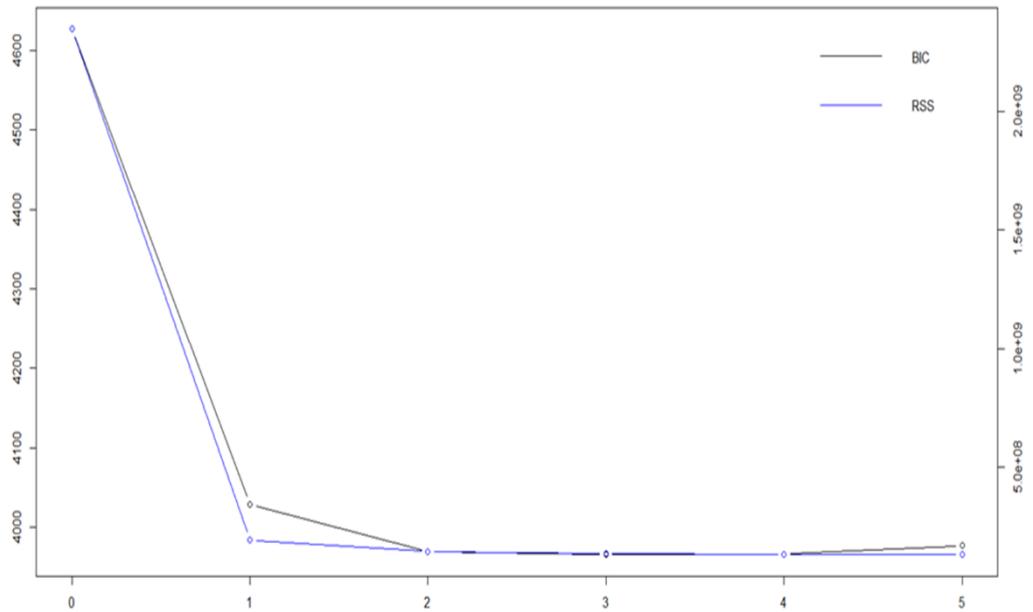

(b)

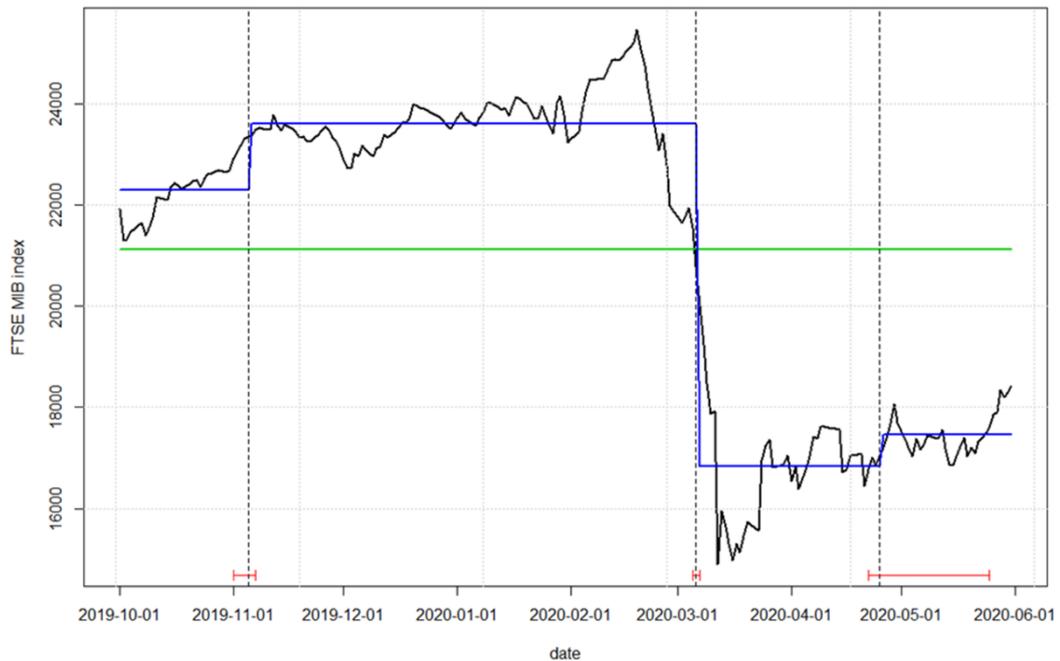

Figure 5. (a) BIC and Residual Sum of Squares using the FTSE-MIB index values; (b) Breakpoints in FTSE-MIB index values

We also investigate about a possible association between sentiment scores and stock exchange index values. Since we observe a time lag between the main breakpoint in sentiment scores and the corresponding breakpoint in FTSE-MIB values, we hypothesize that this association might be present even when applying a time lag (i.e. past sentiment scores are associated with stock exchange index values in the next days). In this respect, we estimate the regression model $y_t = \alpha^{(k)} + \beta^{(k)} x_{t-k} + \varepsilon_t$, where $y_t$ is the price of FTSE-MIB observed at day $t$; $x_{t-k}$ is the sentiment score observed at time $t-k$ and $k$ indicates the time lag. To evaluate the strenght of the association between $x_{t-k}$ and $y_t$ at different time lags, we estimate separate models for $k$ ranging from zero days (no lag, comparing sentiment scores and stock exchange values for the same day) to 50 days (sentiment scores from one day are compared with stock exchange values of 50 days later). Furthermore, to evaluate if the association between sentiment scores and FTSE-MIB prices depends from the specificity of the method used to perform sentiment analysis, we estimate the model separately for each method and each lexicon used to compute sentiment scores. We are mainly intersted in observing how the sign and the magnitude of the estimated regression coefficient $\hat{\beta}^{(k)}$ varies respect to the different method/lexicon used for sentiment analysis and the different time lag. In view of that, we concentrate on the standardized value of $\hat{\beta}^{(k)}$ obtained for each of the 306 estimated models (6 methods/lexicons and 51 time lags).

Results obtained for standardized $\hat{\beta}^{(k)}$ are represented in Figure 6 whilst the values of the standardized estimated coefficient together with the $R^2$ are reported in Supplementary Table 1. The standardized estimated values of $\beta^{(k)}$ at lag zero indicates a strong association between sentiment scores and FTSE-MIB values at lag zero ( $\hat{\beta}^{(k)} \geq 0.70$, p < 0.001, for all methods/lexicons). As showed in Figure 6, a strong association is observed also when considering a lag up to 10-15 days, whilst the same association decreases when the lag exceeds two weeks. Importantly, all the values of $\hat{\beta}^{(k)}$ are postive and significant (p < 0.001) for all the lags and no important differences are observed when changing the method/lexicon used for sentiment analysis. Overall, these findings enforce the idea that a change in sentiment scores can be considered as an early detection signal (up to two weeks earlier) for potential effects on the stock market values.

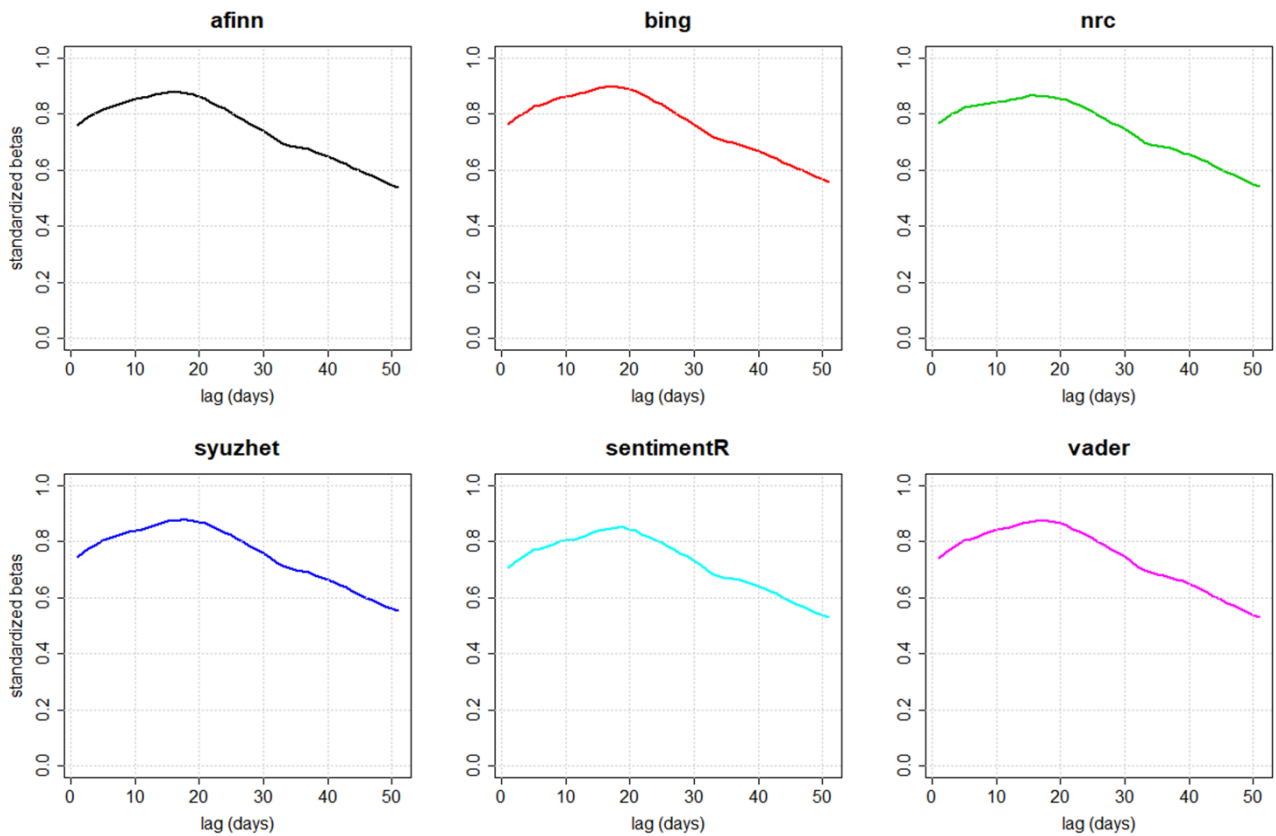

Figure 6. Standardized betas for linear regressions with FTSE-MIB index values as response variable and sentiment scores as predictor (from no lag up to 50 days lag)

## 5. Evaluation of tweets' sentiment using machine learning classifiers

Finally, we investigate about polarity of tweets and compare the performances of two machine learning classifiers widely used in the analysis of Twitter's data, i.e. NB and SVM, in the prediction of tweet polarity before and after the COVID-19 outbreak. Considering that, after the outbreak, the topic was largely discussed on Twitter, we aim to evaluate whether the two classifiers show differences in performance when analyzing texts pertaining to more diverse topics (whole dataset and Period A) or a more homogeneous topic (Period B). To this aim, we randomly sampled 16,128 tweets (9,444 from Period A and 6,684 from Period B, respectively) and manually labeled them into "positive", "negative" or "neutral" according to their specific content. Of these, 3,623 tweets are labeled as positive or negative (1,874 from Period A and 1,749 from Period B) and are used for the subsequent analysis. These tweets are first preprocessed (removal of punctuation, numbers and stop words) and used to form a document text matrix using the tm R package [9]. For both machine learning classifiers, the dataset of 3,623 positive and negative tweets, as well as the two subsets pertaining to Period A and B, were randomly split into a training set including 80% of cases and a test set including 20% of cases. Naïve Bayes and Support Vector Machines are used to train the model on the training set observations and to validate their accuracy on the test set observations. As the specific content of the tweets is more related to Covid-19 in Period B, we expect the two classifiers to perform better for tweets appearing in this period compared to those appearing in the pre-outbreak period (Period A).

**5.1 Naïve Bayes**

Naïve Bayes (NB) classifiers are supervised learning algorithms used for more than two decades for different classification problems, including spam filtering and text classification [32]. A NB classifier is based on the Bayes theorem which states that, given two events A and B, we can compute the conditional probability of A given B as the probability of B given A times the probability of A, all divided by the probability of B. Notationally, we have:

$$P(A|B) = \frac{P(B|A)\, P(A)}{P(B)} \tag{1}$$

A NB classifier assumes that features $x_j$ $(j = 1, ...., p)$ are conditionally independent:

$$P(x_j|X = x_1, ..., x_p) = P(x_j) \quad \forall j \in [1, p]$$

Thus, considering a categorical response variable $y$ with $k$ classes $(k \geq 2)$ the Bayes theorem is reformulated as:

$$P(C = C_k \mid X = x_1, ..., x_p) = \frac{P(C_k) \prod_{j=1}^{p} P(x_j \mid C_k)}{P(x_1, ..., x_p)} \tag{2}$$

where $C$ is one of the $C_k$ classes of $y$, and $X$ is a vector of random variables $x_j$. The task is thus to assign each tweet $t_i (i = 1, ...., n)$ to one of the $k$ classes of $y$. In our case, tweets can be classified either in positive or negative ($k = 2$). Since the denominator in Eq. 2 is the same for each class, we can get rid of it and approximate Eq. 2 as follows:

$$P(C_k \mid x_1, ..., x_p) \propto P(C_k) \prod_{i=1}^{p} P(x_j \mid C_k) \tag{3}$$

so that the assigned class $\hat{C}$ is chosen according to:

$$\hat{C} = \text{argmax}\; P(C_k) \prod_{i=1}^{p} P(x_j \mid C_k) \tag{4}$$

Therefore, any tweet $t_i$ is classified into one, and only one, of the $C_k$ classes.

**5.2 Support Vector Machine (SVM)**

SVM is a non-probabilistic binary linear classifier [36], as it is not based on a specific probability distribution. In a linear setting, SVM detects a line, so called hyperplane, which best separates data. As in the case of NB, the task is to assign each tweet $t_i (i = 1, ...., n)$ to one of the two classes (positive or negative) of the categorical response $y$. The hyperplane that best splits the data is that positioned as far as possible from the nearest points of the two classes. The distance between a point and the hyperplane is called margin. The larger these margins (hard margins), the smaller the probability of misclassification error. To train a linear SVM, we consider data $(\vec{x}_1, y_1), ..., (\vec{x}_n, y_n)$, where $\vec{x}_i$ are vectors of feature values of observation $i$ ($i = 1, ...., n$), and $y_i$ are labels (+1 or –1, depending on which class an observation belongs to). The goal is to find a hyperplane

$$\vec{w} \cdot \vec{x}_i + b = 0 \tag{5}$$

that best separates the two classes ($\vec{w}$ is the normal vector to the hyperplane). It is possible to add constraints about the margin to avoid having any observation within the margins. These constraints are usually specified as: $y_i(\vec{w} \cdot \vec{x}_i + b) \geq 1$, for all $1 \leq i \leq n$.

One of the main problems that may arise when looking for the best hyper-plane is that even a single new observation might cause a great shift of the hyper-plane, reducing the margin by a great amount. In such cases, it is preferable to have a classifier able to tolerate a sub-optimal separation of the two classes but apt to deliver a higher overall classification performance (soft margins). In practice, some points are allowed to be on the wrong side of the margin, or even of the hyper-plane. Since each margin is distant $1/\|\vec{w}\|$ from the hyper-plane that separates the classes, the distance between the two margins is equal to $2/\|\vec{w}\|$. Thus, this distance is maximized when $\|\vec{w}\|$ is minimized. To achieve this, the following a quadratic programming problem needs to be solved:

$$\min f : \frac{1}{2} \|w\|^2 \tag{6}$$
$$\text{s. t. } g: y_i(\vec{w} \cdot \vec{x}_i + b) \geq 1$$

Eq.6 is a constrained optimization problem that can be solved using the method of Lagrange multipliers. It requires a so-called slack variable $\xi_i$ such that the constraint can be expressed as

$$y_i\left(\vec{w} \cdot \vec{x}_i + b\right) \geq B(1 - \xi_i) \tag{7}$$

where B = $1/\|\vec{w}\|$, $\xi_i \geq 0$ and $\sum_{i=1}^n \xi_i \leq K$. Therefore, some points are allowed to be on the wrong side, but only up to a given distance less or equal to the constant $K$. As a result, the optimization problem specified in Eq 6, subject to the constraint specified in Eq 7, becomes

$$\begin{cases} y_i\left(\vec{w} \cdot \vec{x}_i + b\right) \geq B(1 - \xi_i) \\ \min \|w\|^2 \text{ s. t. } \xi_i \geq 0 \ \sum_{i=1}^n \xi_i \leq K \end{cases} \tag{8}$$

### 5.3 Comparison between the performance of naïve Bayes and Support Vector Machine

NB and SVM are estimated using the R package e1071 [22]. Results about their test-set performance are reported in Table 2.

Respect to the whole test set (Period A + Period B), NB reaches a good accuracy and high precision for both positive and negative tweets (precision is 0.87 and 0.78, respectively). Restricting the analysis to Period A, the classifier maintains a good accuracy and still shows a good performance in the classification of negative tweets. However, performance in the classification of positive tweets get worsen. An opposite result is observed when restricting on Period B. Therefore, NB does not allow us to observe an improvement in classification accuracy when analyzing a dataset in which a specific topic, i.e. the COVID-19 pandemic, characterizes the content of a larger part of negative tweets (Period B) compared to periods in which more diverse topics are discussed by users (whole dataset and Period A).

**Table 2. Performance of naïve Bayes and Support Vector Machine in the classification of tweets**

|  | Naïve Bayes | | Support Vector Machine | |
|---|---|---|---|---|
|  | Positive | Negative | Positive | Negative |
| *Whole dataset* | | | | |
| Precision | 0.87 | 0.78 | 0.83 | 0.78 |
| Recall | 0.75 | 0.89 | 0.76 | 0.84 |
| F1-score | 0.81 | 0.83 | 0.79 | 0.81 |
| Support | 364 | 360 | 364 | 360 |
| Accuracy | | 0.82 | | 0.80 |
| *Period A* | | | | |
| Precision | 0.74 | 0.81 | 0.66 | 0.77 |
| Recall | 0.60 | 0.89 | 0.52 | 0.85 |
| F1-score | 0.66 | 0.84 | 0.58 | 0.81 |
| Support | 212 | 113 | 212 | 113 |
| Accuracy | | 0.79 | | 0.74 |
| *Period B* | | | | |
| Precision | 0.84 | 0.61 | 0.88 | 0.60 |
| Recall | 0.80 | 0.86 | 0.83 | 0.70 |
| F1-score | 0.86 | 0.71 | 0.86 | 0.64 |
| Support | 93 | 257 | 93 | 257 |
| Accuracy | | 0.81 | | 0.79 |

Next, SVM with a linear kernel is estimated in the same way as NB. Again, when analyzing the whole dataset (Period A + Period B) the classifier shows a good performance according to all metrics, with slightly lower precision but higher recall and F1-score for negative compared to positive tweets. When analyzing Period A, consistent with the results obtained for NB, SVM shows lower accuracy and worse performance in the classification of positive tweets, while maintaining good precision, recall and F1-score in the classification of negative tweets. The opposite result is observed when analyzing Period B.

Results obtained for the whole dataset (Period A + B, Table 2) evidence that the two classifiers show similar accuracy and F1 score, but more pronounced differences are observed in terms of precision and recall, with NB showing a slightly better performance for most metrics. While both classifiers show higher precision in the classification of positive compared to negative tweets, the opposite result is observed for recall and F1-score. Results obtained for the tweets collected during Period A reveal that, for all metrics, both classifiers show a better performance in the classification of negative compared to positive tweets. When analyzing the tweets collected during Period B, both classifiers show better precision and F1-score in the classification of positive compared to negative tweets. Conversely, recall is higher for negative tweets using NB and for positive tweets using SVM. Overall, these results are consistent with our initial assumption that classifiers' performance is improving following the beginning of the outbreak as the content of tweets is more specifically focused on the pandemic. In conclusion, although the two classifiers do not show differences as regard to accuracy, NB would be preferable due to its overall better performance in terms of precision for both positive and negative tweets.

## 6. Concluding remarks

In this paper, we analyze the sentiment towards Italy before and after the COVID-19 outbreak using lexicon-based and machine learning classifiers applied to real data collected from Twitter. We observe a substantial rise in negative emotions towards Italy in correspondence of the first Italian case of COVID-19 followed by a change towards more neutral or slightly positive values starting two months later. Besides being useful to interpret the general sentiment towards a country as a proxy of the perceived country reputation, we find that sentiment scores can be also used to early detect changes in stock exchange values. Future research is addressed to assess the sentiment towards different countries, to verify whether similar findings might be observed also in cases in which the outbreak developed with different rates of escalation and/or in the cases when severity is managed by the governments with different alternative or concomitant measures such as social distancing, lockdown or travel restrictions.

The results of this research must be interpreted in the context of its limitations. First, data was collected from a single social network (i.e., Twitter). It is possible that the results could vary in the case many social networks are considered. However, as shown in the literature review, Twitter is widely used to evaluate reactions to important events due to its diffusion and ease of use. As already mentioned, future work should also focus on other countries, as our findings could vary due to the different country's reputation or other factors (e.g. cultural or socioeconomic factors). Despite these limitations, we believe that this analysis is helpful to understand how the sentiment towards Italy, one of the first countries severely affected by COVID-19, evolved through time. Furthermore, it can also help to shed light on the relationship between country reputation and the possible economic repercussions of an event of this magnitude.

## Declarations

Funding: none

Conflicts of interest/Competing interests: none

Availability of data and material: data available upon request

Code availability: code available upon request

**Supplementary Table 1. Linear regression models with FTSE-MIB index values as response and sentiment scores as predictor**

| Lag | afinn $\hat{\beta}^{(k)}$ | $R^2$ | bing $\hat{\beta}^{(k)}$ | $R^2$ | nrc $\hat{\beta}^{(k)}$ | $R^2$ | syuzhet $\hat{\beta}^{(k)}$ | $R^2$ | sentimentR $\hat{\beta}^{(k)}$ | $R^2$ | vader $\hat{\beta}^{(k)}$ | $R^2$ |
|---|---|---|---|---|---|---|---|---|---|---|---|---|
| 0 | 0.76 | 0.58 | 0.76 | 0.58 | 0.77 | 0.59 | 0.75 | 0.50 | 0.71 | 0.55 | 0.74 | 0.55 |
| 1 | 0.78 | 0.60 | 0.78 | 0.61 | 0.78 | 0.61 | 0.76 | 0.58 | 0.73 | 0.52 | 0.76 | 0.58 |
| 2 | 0.79 | 0.63 | 0.80 | 0.64 | 0.80 | 0.64 | 0.78 | 0.61 | 0.74 | 0.55 | 0.78 | 0.60 |
| 3 | 0.80 | 0.65 | 0.81 | 0.66 | 0.81 | 0.66 | 0.79 | 0.63 | 0.75 | 0.57 | 0.79 | 0.62 |
| 4 | 0.82 | 0.67 | 0.83 | 0.68 | 0.82 | 0.68 | 0.81 | 0.65 | 0.77 | 0.59 | 0.81 | 0.65 |
| 5 | 0.82 | 0.68 | 0.83 | 0.69 | 0.83 | 0.68 | 0.81 | 0.66 | 0.77 | 0.60 | 0.81 | 0.66 |
| 6 | 0.83 | 0.69 | 0.84 | 0.70 | 0.83 | 0.69 | 0.82 | 0.67 | 0.78 | 0.61 | 0.82 | 0.67 |
| 7 | 0.84 | 0.70 | 0.85 | 0.72 | 0.84 | 0.70 | 0.83 | 0.68 | 0.79 | 0.62 | 0.83 | 0.68 |
| 8 | 0.85 | 0.72 | 0.86 | 0.73 | 0.84 | 0.71 | 0.84 | 0.70 | 0.80 | 0.64 | 0.84 | 0.70 |
| 9 | 0.85 | 0.73 | 0.86 | 0.74 | 0.84 | 0.71 | 0.84 | 0.71 | 0.80 | 0.65 | 0.84 | 0.71 |
| 10 | 0.86 | 0.74 | 0.87 | 0.75 | 0.84 | 0.71 | 0.84 | 0.71 | 0.81 | 0.65 | 0.85 | 0.72 |
| 11 | 0.86 | 0.75 | 0.87 | 0.76 | 0.85 | 0.72 | 0.85 | 0.72 | 0.81 | 0.66 | 0.85 | 0.73 |
| 12 | 0.87 | 0.75 | 0.88 | 0.77 | 0.85 | 0.73 | 0.86 | 0.73 | 0.82 | 0.67 | 0.86 | 0.74 |
| 13 | 0.87 | 0.76 | 0.89 | 0.79 | 0.86 | 0.74 | 0.87 | 0.75 | 0.83 | 0.69 | 0.87 | 0.75 |
| 14 | 0.88 | 0.77 | 0.89 | 0.80 | 0.87 | 0.75 | 0.87 | 0.76 | 0.84 | 0.70 | 0.87 | 0.76 |
| 15 | 0.88 | 0.77 | 0.90 | 0.80 | 0.87 | 0.75 | 0.88 | 0.77 | 0.84 | 0.71 | 0.87 | 0.76 |
| 16 | 0.88 | 0.77 | 0.90 | 0.81 | 0.86 | 0.75 | 0.88 | 0.77 | 0.85 | 0.72 | 0.87 | 0.76 |
| 17 | 0.87 | 0.76 | 0.90 | 0.81 | 0.86 | 0.74 | 0.88 | 0.77 | 0.85 | 0.72 | 0.87 | 0.76 |
| 18 | 0.87 | 0.76 | 0.89 | 0.80 | 0.86 | 0.74 | 0.88 | 0.76 | 0.85 | 0.72 | 0.87 | 0.76 |
| 19 | 0.86 | 0.75 | 0.89 | 0.79 | 0.86 | 0.73 | 0.87 | 0.76 | 0.85 | 0.71 | 0.86 | 0.75 |
| 20 | 0.86 | 0.73 | 0.88 | 0.78 | 0.85 | 0.72 | 0.86 | 0.75 | 0.84 | 0.70 | 0.86 | 0.74 |
| 21 | 0.84 | 0.70 | 0.87 | 0.75 | 0.84 | 0.70 | 0.85 | 0.73 | 0.83 | 0.68 | 0.85 | 0.71 |
| 22 | 0.83 | 0.69 | 0.86 | 0.73 | 0.83 | 0.69 | 0.84 | 0.71 | 0.82 | 0.66 | 0.84 | 0.70 |
| 23 | 0.82 | 0.67 | 0.85 | 0.71 | 0.82 | 0.67 | 0.83 | 0.69 | 0.81 | 0.65 | 0.83 | 0.68 |
| 24 | 0.81 | 0.65 | 0.83 | 0.69 | 0.81 | 0.66 | 0.82 | 0.68 | 0.80 | 0.63 | 0.81 | 0.66 |
| 25 | 0.79 | 0.62 | 0.82 | 0.67 | 0.80 | 0.63 | 0.81 | 0.65 | 0.78 | 0.61 | 0.80 | 0.64 |
| 26 | 0.78 | 0.60 | 0.81 | 0.65 | 0.78 | 0.61 | 0.80 | 0.63 | 0.77 | 0.59 | 0.79 | 0.62 |
| 27 | 0.76 | 0.58 | 0.79 | 0.62 | 0.77 | 0.59 | 0.78 | 0.61 | 0.76 | 0.57 | 0.77 | 0.59 |
| 28 | 0.75 | 0.57 | 0.78 | 0.60 | 0.76 | 0.58 | 0.77 | 0.59 | 0.75 | 0.56 | 0.76 | 0.58 |
| 29 | 0.74 | 0.55 | 0.76 | 0.58 | 0.75 | 0.56 | 0.76 | 0.58 | 0.73 | 0.54 | 0.75 | 0.56 |
| 30 | 0.73 | 0.52 | 0.75 | 0.56 | 0.73 | 0.53 | 0.74 | 0.55 | 0.72 | 0.51 | 0.73 | 0.53 |
| 31 | 0.71 | 0.50 | 0.73 | 0.53 | 0.72 | 0.51 | 0.72 | 0.52 | 0.70 | 0.48 | 0.71 | 0.50 |
| 32 | 0.70 | 0.48 | 0.72 | 0.51 | 0.70 | 0.49 | 0.71 | 0.51 | 0.68 | 0.47 | 0.70 | 0.49 |
| 33 | 0.69 | 0.47 | 0.71 | 0.50 | 0.69 | 0.48 | 0.70 | 0.49 | 0.68 | 0.45 | 0.69 | 0.47 |
| 34 | 0.68 | 0.46 | 0.70 | 0.49 | 0.69 | 0.47 | 0.70 | 0.49 | 0.67 | 0.44 | 0.68 | 0.46 |
| 35 | 0.68 | 0.46 | 0.70 | 0.48 | 0.68 | 0.47 | 0.69 | 0.48 | 0.67 | 0.44 | 0.68 | 0.46 |
| 36 | 0.67 | 0.45 | 0.69 | 0.48 | 0.68 | 0.46 | 0.69 | 0.47 | 0.66 | 0.44 | 0.67 | 0.45 |
| 37 | 0.67 | 0.44 | 0.68 | 0.46 | 0.67 | 0.45 | 0.68 | 0.46 | 0.66 | 0.43 | 0.67 | 0.44 |
| 38 | 0.66 | 0.43 | 0.67 | 0.45 | 0.66 | 0.44 | 0.67 | 0.45 | 0.65 | 0.42 | 0.66 | 0.43 |
| 39 | 0.65 | 0.42 | 0.67 | 0.44 | 0.66 | 0.43 | 0.67 | 0.44 | 0.64 | 0.41 | 0.65 | 0.42 |
| 40 | 0.64 | 0.41 | 0.66 | 0.43 | 0.65 | 0.42 | 0.66 | 0.43 | 0.63 | 0.40 | 0.64 | 0.41 |
| 41 | 0.63 | 0.40 | 0.65 | 0.42 | 0.64 | 0.41 | 0.65 | 0.42 | 0.62 | 0.38 | 0.63 | 0.39 |
| 42 | 0.62 | 0.39 | 0.64 | 0.41 | 0.63 | 0.39 | 0.64 | 0.40 | 0.61 | 0.37 | 0.62 | 0.38 |
| 43 | 0.61 | 0.37 | 0.63 | 0.39 | 0.62 | 0.38 | 0.62 | 0.38 | 0.60 | 0.36 | 0.61 | 0.36 |
| 44 | 0.60 | 0.36 | 0.62 | 0.38 | 0.60 | 0.36 | 0.61 | 0.37 | 0.59 | 0.34 | 0.60 | 0.35 |
| 45 | 0.59 | 0.34 | 0.61 | 0.37 | 0.59 | 0.35 | 0.60 | 0.36 | 0.58 | 0.33 | 0.58 | 0.34 |
| 46 | 0.58 | 0.33 | 0.60 | 0.36 | 0.58 | 0.34 | 0.59 | 0.35 | 0.57 | 0.32 | 0.57 | 0.32 |
| 47 | 0.57 | 0.32 | 0.59 | 0.34 | 0.57 | 0.32 | 0.58 | 0.33 | 0.56 | 0.31 | 0.56 | 0.31 |
| 48 | 0.56 | 0.31 | 0.58 | 0.33 | 0.56 | 0.31 | 0.57 | 0.32 | 0.55 | 0.29 | 0.55 | 0.30 |
| 49 | 0.55 | 0.30 | 0.57 | 0.32 | 0.55 | 0.30 | 0.56 | 0.31 | 0.54 | 0.29 | 0.54 | 0.29 |
| 50 | 0.54 | 0.29 | 0.56 | 0.31 | 0.54 | 0.29 | 0.55 | 0.30 | 0.53 | 0.28 | 0.53 | 0.28 |

The table reports standardized betas ($p < 0.001$) and model $R^2$ for a linear regression with FTSE-MIB index values as response variable and sentiment scores as predictor. Different time lags, from zero days (no lag, comparing sentiment scores and stock exchange values for the same day) to 50 days (sentiment scores from one day are compared with stock exchange values of 50 days later) have been used.